# Efficient Local Density Estimation Strategy for VANETs

Haouari Noureddine
USTHB
Algiers, Algeria
nhaouari@usthb.dz

Moussaoui Samira
USTHB
Algiers, Algeria
smoussaoui@usthb.dz

*Abstract*— **Local vehicle density estimation is increasingly becoming an essential factor of many vehicular ad-hoc network applications such as congestion control and traffic state estimation. This estimation is used to get an approximate number of neighbors within the transmission range since beacons do not give accurate accuracy about neighborhood. These is due to the special characteristics of VANETs such as high mobility, high density variation. To enhance the performance of these applications, an accurate estimation of the local density with minimum of overhead is needed. Most of the proposed strategies address the global traffic density estimation without a big attention on the local density estimation. This paper proposes an improved approach for local density estimation in VANETs in terms of accuracy and overhead. The simulation results showed that our strategy allows an interesting precision of estimation with acceptable overhead.**

*Keywords-VANETs; density estimation; local density;*

## I. INTRODUCTION

Vehicle density estimation schemes are the key element to ensure better performance of many VANETs (Vehicular Ad-hoc Networks) applications like: congestion control and traffic density estimation. We consider the local density as the total number of the vehicles within the transmission range where vehicles might not be able to send successfully packets. The range where the transmitting vehicles are able to decode packets is the communication range. The performance of such applications is highly dependent on the accuracy of the local density estimation. For instance, in congestion control, the local density can be used to detect the network congestion by the estimation of the generated load on the control channel. This can be calculated by knowing the total number of neighbors and the estimated generated load per node. Also, in traffic estimation, local vehicle density estimation is widely used to estimate the global density based on the estimations provided by the different vehicles. Thus, the performance of the local density estimation has a significant impact on many applications in VANETs.

According to [1], in situations where there is a high message load the reliable transmission range is reduced by up to 90%. This degradation causes low neighborhood awareness. As a result, the estimation of the density through beacons will give the number of the neighbors in communication range not in the transmission range. Thus, such degradation causes a very limited view on the neighborhood which might perturb the good functioning of many applications that use beacons to estimate the density.

Due to the lack of using beacons, many other density estimation approaches were proposed in the literature. Some estimation strategies are based on the speed of the vehicle itself to estimate the density on the road. These approaches have the advantage of zero extra overhead. However, the accuracy of these strategies is obviously not guaranteed since the speed does not always reflect the density. For instance, a vehicle might decrease its speed at intersections without having a high density on the road.

Other density estimation strategies use message exchange between vehicles to enhance the density precision. The density is considered to be more accurate by the cooperation of all vehicles. D-FPAV strategy [2], which is a congestion control strategy, uses extended beacons to exchange the information neighbors. These extended beacons are normal beacons but has extra information about neighborhood. These messages are sent every n sent beacon. D-FPAV density estimation approach gives an interesting precision. However, this approach generates high overhead on the control channel. DVDE/SPAV [3] strategy is proposed to overcome this problem by introducing a lower quality of density with less overhead. DVDE gives very interesting results comparing with D-FPAV in terms of accuracy and overhead. Despite DVDE decreases the generated overhead, it still has an extra overhead that could be avoided. In this paper, we propose a local density estimation approach with higher accuracy and with less overhead by improving DVDE strategy.

The paper is organized into four sections as follows: Section 2 presents related work about local density estimation. Section 3 introduces ELDES strategy. The simulation and the evaluation of ELDES are presented in Section 4. Section 5 concludes the paper with outlooks on the future work.

## II. RELATED WORK

There have been few works addressing the local density estimation in spite of its numerous applications in VANETs.

Most of the works address traffic density estimation which aims to calculate the density of a specific road or section. In this paper, we are interested in the strategies used to estimate the local density. We do not consider approaches that use special infrastructure such as inductive loop detectors or traffic surveillance cameras. More specifically, we review the free infrastructure solutions where the only options for vehicles are using communication or observing movement. The reviewed strategies can be divided into two categories: speed-based strategies [4-7] and communication-based strategies [2, 3, 8, 9, 10].

Speed-based mechanisms are based on speed–density relationships to estimate the density on the road. In [4], the estimation of density is based on vehicles mobility patterns that are car-following model and two-fluid model. The density estimation is used to dynamically choose the transmission range. This approach could not estimate the density in free-flow traffic due to the absence of interactions between vehicles. The authors in [5] propose Velocity Aware Density Estimation (VADE). In VADE, each single vehicle tracks its own velocity and acceleration. The traffic density is estimated based on the observed speed and the traffic flow theory [6]. For instance, if the vehicles move with high speed, the density would be estimated as sparse. In [7], fluid dynamics and car-following model are used to estimate the vehicle density. Nevertheless, these strategies could give inaccurate results about the density since the speed is not always related to the density of vehicles on the road. For example, vehicles could stop at intersections without having high density. Moreover, these strategies are developed for the global density estimation not for the local density estimation where the precision is relatively high.

Communication-based strategies are based on exchanging messages between vehicles to estimate the density.

In [8], the density is calculated based on the number of the local neighbors. Then, the global density of the road is concluded based on supposing that the inter-vehicle spacing is exponentially distributed which might not be the case for all possible traffic scenarios.

In [9], the proposed strategy calculates the density of a specific target area. We are just interested in how it calculates the local density of a specific vehicle. Using a vehicle called "sampler", a message called "POLL" is broadcasted at each sampling instant that contains the position and the radius of the target area. When a vehicle in the target area receives this message, it will reply after waiting for random time to avoid flooding the receiver. After a period of time, the sampler will count the received messages to estimate the density. This strategy gives an accurate results comparing with the actual density. However, this strategy needs long time relatively and it cannot be used for critical applications like congestion control where timely reaction is very critical.

In [10], the authors adapted mechanisms from system size estimation in P2P for vehicular density estimation. The authors propose three distributed and free-infrastructure mechanisms, namely, Sample & Collide, Hop Sampling and Gossip-based Aggregation for VANETs. The simulation results show the high performance of Hop Sampling. In this approach, the vehicle wants to know the density starts by sending "gossip message" to the neighbors. Each vehicle receives this particular message, it will update its hopcount if this is the first received message or the received value is less than the current value. After the update, the vehicle forwards this message with (hopcount+1). Then, any vehicle receives the "gossip message" will send a reply with certain probability. If hopcount is less than a specific threshold called "minHopsReporting", the probability is one. Otherwise, the probability decreases as far as the vehicle position. Their simulation results show the high accuracy of the proposed mechanism. However, this mechanism could lead to wrong results due to the existence of some vehicles that do not reply because of their low probability. Moreover, it is difficult to specify the target area to be as the local transmission range of a specific vehicle. Also, the delay is another problem for this mechanism which affects the critical applications.

In [2], the authors propose a congestion control protocol called D-FPAV. In the proposed strategy, they need to find out the total number of all neighbors in the transmission range (not only within communication range). For that, they propose to use multi-hop strategy where a vehicle sends a piggybacked beacon each n beacon containing its neighbors. When a vehicle receives these extended beacons, it will be aware of all the vehicles in its transmission range. This information is used later to estimate the load on the channel. This approach suffers from the high generated overhead in the channel.

In [3], the authors address the high overhead problem of D-FPAV. They propose their own density estimation strategy DVDE to overcome the overhead generated by the extended beacons. Their approach based on the segmentation of the transmission range into an odd number of segments and then instead of sending neighbors in extended beacons, the vehicles send the density of each segment every n beacon. When a node receives the density of segments, it chooses the nearest to the center of the target segment. If the segments are not the same, it uses linear interpolation to estimate the density. By gathering this information from different vehicles, the vehicle could enhance the accuracy of its estimation. This approach reduces the overhead comparing to the D-FPAV method. However, Even that DVDE strategy has given interesting results in terms of accuracy and overhead, it still has some shortcomings. In fact, the process of data density sharing between vehicles is difficult due to the different segment positions as DVDE strategy supposes that each vehicle has its own segments. For this reason, in DVDE strategy, the authors propose to use linear interpolation to estimate the density of a target segment even if the segments are different which happens in most cases. However, this approach could give less accurate results if the vehicles are not uniformly distributed. Moreover, the shared information of periodic extended beacons could be useless if the vehicles are in the same area sharing the same information. This periodic redundancy creates an extra overhead could be avoided if only some selected vehicles share their information. Our work is based on improving the DVDE in order to develop a local density estimation strategy more accurate with lower overhead.

III. EFFICIENT LOCAL DENSITY ESTIMATION STRATEGY (ELDES)

Most of the discussed approaches share the common goal of enabling higher density estimation efficiency. However, there have been few contributions addressed the local density estimation. In this section, we present ELDES as an improvement of DVDE strategy. The primary goal of ELDES is to estimate the local density within vehicle transmission range with higher accuracy and less overhead.

In designing ELDES, the following assumptions are made:

- Vehicles are equipped with omnidirectional antennas.
- All vehicles have the same receiving sensitivity and similar transmission ranges.
- Each vehicle is aware of its geographical location and velocity through a global Positioning system (GPS) device.
- Roads are segmented into zones; each zone is identified. All vehicles could determine the zone where they are bases on preloaded digital.

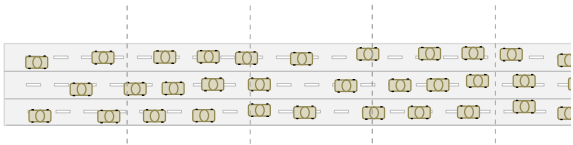

Figure 1.  A segmented road

The main drawbacks of DVDE strategy is the redundancy of extended beacons and the less accuracy of estimation by using linear interpolation. To overcome these shortcomings, in ELDES, the vehicles are supposed to be in segmented roads (Figure 1) where each vehicle could identify each segment on the road. When a vehicle passes on a centre of a segment, it sends an extended beacon if it did not receive any extended beacon from this position in the last **ΔT.** The extended beacons are build based on the received normal beacons and extended beacons, therefore if a vehicle did not receive an extended beacon for specific segment, it estimates its value based on the received normal beacons. For each segment, ELDES looks for the nearest vehicle to it, and it extracts the segment value from the vehicle data, if it is not outdated. Therefore, ELDES increases the accuracy of the density estimation by avoiding the using of linear interpolation since all the nodes have the same segments. Also, by using fixed segments, only the nodes in the centre of the segments share their information which lessens the overhead on the channel. This overhead is decreased more by avoiding sending extended beacons before a period of time **ΔT** from the same segment. This is to avoid the redundancy of information if the vehicles are near to the same centre of a specific segment.

The following algorithm shows how ELDES works:

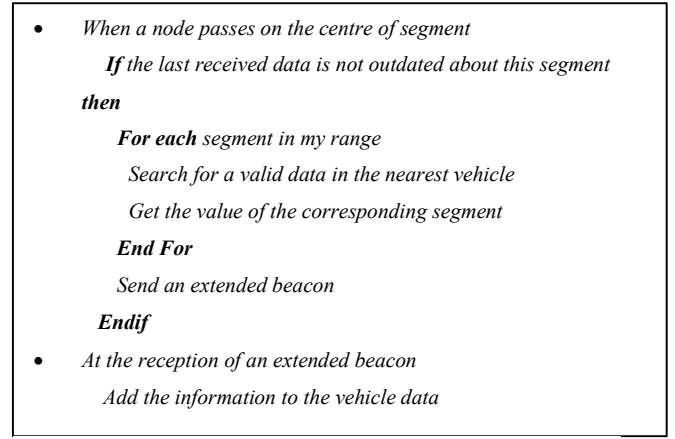

- *When a node passes on the centre of segment*
    *If the last received data is not outdated about this segment*
  **then**
    **For each** *segment in my range*
      *Search for a valid data in the nearest vehicle*
      *Get the value of the corresponding segment*
    **End For**
    *Send an extended beacon*
  **Endif**
- *At the reception of an extended beacon*
    *Add the information to the vehicle data*

Figure 2.  The Functioning of ELDES

In the next section we evaluate the performance of ELDES and we compare it with DVDE strategy.

IV. SIMULATION AND EVALUATION

*A. Simulation Environment*

In this section, we present our VANET simulation setup. Simulation results presented in this paper were obtained using the NS-2 simulator [11]. The NS-2 is a discrete event simulator developed at the University of California. We choose NS-2 for its credibility among network research community. Also, to have more realistic results we have used an overhauled MAC/PHY-model [12] adapted to the characteristics of IEEE 802.11P (the standard of the inter-vehicle communications). Table I presents the simulation parameters of the medium access and physical layer according to the IEEE 802.11P standard. The simulations are run using the deterministic Two-Ray Ground propagation model.

TABLE I.   MEDIUM ACCESS AND PHYSICAL LAYER CONFIGURATION PARAMETERS FOR IEEE 802.11P

| Parameter | Value |
| --- | --- |
| Frequency | 5.9 GHz |
| Data rate | 6 Mbps |
| Carrier Sense Threshold | -96 dBm |
| Noise floor | -99 dBm |
| SINR for preamble capture | 4 dB |
| SINR for frame body capture | 10 dB |
| Slot time | 16 us |
| SIFS time | 32 us |
| Preamble length | 40 us |
| PLCP header length | 8 us |

After careful analysis of various available tools, SUMO [13] was used to generate the movement pattern of vehicles. We use this tool because it is open source, highly portable and can be used to simulate both the microscopic and macroscopic environments. We simulate scenario of 5 km with the following parameters:

TABLE II. NS-2 SIMULATION PARAMETERS

| Parameter | Value |
|---|---|
| MAC | 802.11P |
| Beacon generation | 10 beacons/s |
| Beacon lifetime | 0.3 second |
| Packet size | 500 byte |
| Maximum vehicle velocity | 30 m/s |
| Transmission Range | 1000 m |
| Radio propagation | TwoRayGround |
| Number of vehicles | 160 |

*B. Performance Metrics*

Two performance metrics were used for the evaluation of the performance of ELDES:

**Error ratio:** It is calculated using the following formula:

$$\text{Error ratio} = \frac{|EN - RN|}{RN}$$

Where EN is the number of the estimated neighbors and RN is the number of the real neighbors in the transmission range.

**Overhead:** It is the number of the sent extended beacons. We take the number and not the size because we use the same format of messages in the extended beacons.

*C. Simulation results*

The core algorithm described in Figure 2 is evaluated using simulations. The simulation setup used is described in Table I and Table II. We run the simulation for 10 seconds. At the end of the simulation, each vehicle estimates the local density. The simulation scenarios are made with the same parameters for ELDES and DVDE.

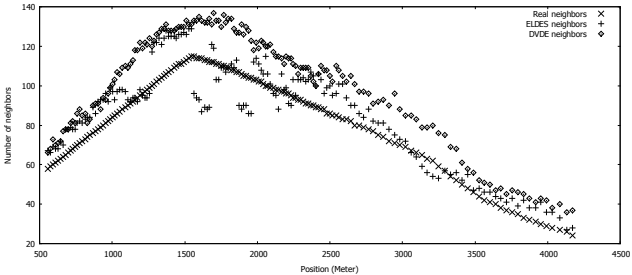

Figure 3. Comparison between ELDES, DVDE and the real neighbors

Figure 3 shows that ELDES is near to the real number of neighbors comparing with DVDE. Despite the high variation of values of ELDES, in most cases the error of ELDES is lower than DVDE. The following table shows the error ratio of both strategies:

TABLE III. COMPARISON OF THE ESTIMATION ERROR BETWEEN ELDES AND DVDE

| | ELDES | DVDE |
|---|---|---|
| **Average error (neighbor)** | +/- 9.43 | +/- 17.33 |
| **Error ratio** | 11.71% | 21.89 % |

TABLE III shows that ELDES outperforms DVDE strategy in terms of accuracy thanks to the using of fixed segments instead of dynamic segments. This makes ELDES more accurate without the using of linear interpolation.

To evaluate the communication overhead generated by the two strategies, we compare the number of the sent extended beacons in both strategies since they have the same format and size.

TABLE IV. COMPARISON OF THE GENERATED OVERHEAD BETWEEN ELDES AND DVDE

| | ELDES | DVDE |
|---|---|---|
| **Number of extended beacons** | 370 | 1436 |

TABLE IV shows that ELDES clearly needs much less overhead than DVDE. This is expected because ELDES avoids the eventual redundancy by properly selecting the vehicles that share their data which are the vehicles on the center of segments. Also, ELDES blocks the sharing of data if an extended beacon was received from the same segment in the last **ΔT** period.

From the previous simulation, it is clear that ELDES can estimate the local density with higher accuracy and with less overhead comparing with DVDE strategy.

V. CONCLUSION

In Vehicular Ad-hoc Networks many applications use a local density estimation strategy to adapt their functioning to the density value. To ensure a high performance for these applications it is very important to have high accurate strategies with low overhead.

We have proposed a segmented-based local density estimation strategy ELDES. This strategy can be used by several VANETs applications like traffic estimation strategies, congestion control protocols and so on.

The proposed strategy ELDES has many advantages. It has less overhead comparing with DVDE strategy. Moreover, it has higher accuracy. Therefore, by using ELDES, the performance of many VANETs applications could be improved.

Future study will address using ELDES as density estimation strategy for a congestion control protocol.